%% file: main.tex
\documentclass{article}

 \usepackage[final]{neurips_2025}

\usepackage[utf8]{inputenc} 
\usepackage[T1]{fontenc}    
\usepackage{hyperref}       
\usepackage{url}            
\usepackage{booktabs}       
\usepackage{amsfonts}       
\usepackage{nicefrac}       
\usepackage{microtype}      
\usepackage{xcolor}         
\usepackage{csquotes}
\usepackage{soul}
\usepackage{comment}
\usepackage[dvipsnames]{xcolor}
 \usepackage{graphicx}
\usepackage{tabularx}
\usepackage{longtable}
\usepackage{array}

\newcommand{\edit}[1]{\textcolor{Black}{#1}}

\newcommand{\commentsenabled}{}

\ifdefined\commentsenabled
\newcommand{\lucy}[1]{\textcolor{blue}{(LQ: #1)}}
\newcommand{\lqedit}[1]{\textcolor{blue}{#1}}
\newcommand{\princessa}[1]{\textcolor{purple}{(Princessa: #1)}}
\newcommand{\elissa}[1]{\textcolor{Plum}{[EMR: #1]}}
\newcommand{\eredit}[1]{\textcolor{Plum}{#1}}
\newcommand{\deepak}[1]{\hl{\textbf{Deepak}: #1}}
\newcommand\dk[1]{\hl{\textbf{DK}: #1}}
\newcommand{\amc}[1]{\textcolor{teal}{AMc: #1}}
\else
\newcommand{\lucy}[1]{}
\newcommand{\lqedit}[1]{}
\newcommand{\princessa}[1]{}
\newcommand{\elissa}[1]{}
\newcommand{\eredit}[1]{}
\newcommand{\deepak}[1]{}
\newcommand{\dk}[1]{}
\newcommand{\amc}[1]{}
\fi

\newcommand{\nPapers}{150}
\newcommand{\nFullyAnnotated}{74}

\title{Stop the Nonconsensual Use of Nude Images in Research}

\author{
Princessa Cintaqia \\
Boston University\\
\texttt{cintaqia@bu.edu}\\
\And
Arshia Arya \\
UC San Diego \\
\texttt{aarshia@ucsd.edu}\\
\And
Elissa M. Redmiles \\
Georgetown University \\
\texttt{elissa.redmiles@georgetown.edu}\\
\And
Deepak Kumar \\
UC San Diego \\
\texttt{kumarde@ucsd.edu}\\
\And
Allison McDonald\thanks{Both authors advised this study.} \\
Boston University \\
\texttt{amcdon@bu.edu}\\
\And
Lucy Qin\footnotemark[1] \\
Georgetown University \\
\texttt{lucy.qin@georgetown.edu}\\
}

\date{May 2025}

\begin{document}

\maketitle

\begin{abstract}
    In order to train, test, and evaluate nudity detection models, machine learning researchers typically rely
    on nude images scraped from the Internet. Our research finds that this content is collected and, in some cases, subsequently \emph{distributed} by researchers without consent, leading to potential misuse and exacerbating harm against the subjects depicted. \textbf{This position paper argues that the distribution of nonconsensually collected nude images by researchers perpetuates image-based sexual abuse and that the machine learning community should stop the nonconsensual use of nude images in research.}
    To characterize the scope and nature of this problem, we conducted a systematic review of papers published in computing venues that collect and use nude images. 
    Our results paint a grim reality: norms around the usage of nude images are sparse, leading to a litany of problematic practices like distributing and publishing nude images with uncensored faces, and intentionally collecting and sharing abusive content. We conclude with a call-to-action for publishing venues and a vision for research in nudity detection that balances user agency with concrete research objectives.

\end{abstract}
    \textbf{Content warning:} this work discusses sexual violence, such as the harms of image-based sexual abuse (Section~\ref{sec:intro}) and the inclusion of sexually violent imagery in datasets (Section~\ref{sec:data-collection}).

\input{sections/1_intro}
\input{sections/2_background}
\input{sections/3_methods}
\input{sections/4_results}
\input{sections/5_discussion}

\begin{ack}

\edit{This work was partially funded by an unrestricted gift from Google. E. M. Redmiles was supported in part by NSF Awards \#2344940 and \#2513313. L. Qin was supported by the Fritz Fellowship through the Initiative for Technology and Society at Georgetown University.}

\end{ack}


\bibliographystyle{plain}
\bibliography{refs/final}
\newpage
\appendix

\input{sections/appendix}

\end{document}

%% file: sections/1_intro.tex
\section{Introduction}
\label{sec:intro}

Nudity detection is a task that has been studied by researchers for decades~\cite{cifuentes2022survey}. For training, testing, and benchmarking nudity detection algorithms, researchers typically scrape images from the Internet or use existing datasets of nude images. While this practice is common for assembling datasets for general image-recognition tasks, the scraping of nude images raises unique ethical concerns. 

Nude images online take different forms. 
Some nude images may have been created and/or made public without the consent of the person depicted. 
Indeed, publicly-accessible forums have been documented to host communities explicitly for the nonconsensual sharing of nude content~\cite{hargreaves2018creep}.
In many other cases, nude images were created \textit{and} posted to a particular website by, or with the consent of, the image subject. However, even images that were consensually shared on publicly-accessible forums (e.g., Reddit) or adult content platforms (e.g., PornHub, OnlyFans) were intended for dissemination to a particular audience under a particular arrangement (e.g., for pay), not for use and/or redistribution in machine learning research.

The nonconsensual creation or distribution of nude images is a form of sexual violence termed image-based sexual abuse (IBSA)~\cite{powell2022perpetration}, which encompasses the nonconsensual creation (e.g., ``upskirting,'' ``downblousing,'' ``deepfakes'') and/or nonconsensual distribution of intimate content, as well as threats to cause these harms~\cite{henry_image-based_2022}. IBSA can lead to serious downstream consequences similar to physical sexual abuse~\cite{mcglynn_its_2021,bates_revenge_2017}. According to some victim-survivors,\footnote{We use the term victim-survivor to capture the range of ways people who have experienced IBSA identify~\cite{sexual_assault_kit_initiative_victim_nodate, williamson_reconsidering_2018, henry_image-based_2022}. We recognize that some who have experienced harm may not identify with ``victim'' or ``survivor''.} one of the most traumatizing aspects of IBSA is that once an image has been distributed online, the subject loses control over how it is further spread and used~\cite{bates_revenge_2017}. 
IBSA is also considered illegal in a growing number of countries (e.g., the United States~\cite{ccri-ncii-map}, Canada~\cite{government_of_british_columbia_intimate_2025}, Mexico~\cite{anastasia_moloney_revenge_2019}, South Korea~\cite{yang2025unveiling}, the United Kingdom~\cite{revenge_porn_helpline_intimate_nodate}).
Recent research has already documented the prevalence of known child sexual abuse material (CSAM) and nonconsensual intimate images of adults in the training datasets of major AI models as a result of unchecked scraping of images from the Internet~\cite{livia_foldes_nsfw_2024, thiel2023identifying, birhane_large_2021, birhane2021multimodal, everest_pipkin_lacework_2020}. 

Through a systematic analysis of 150 research papers on nudity detection, safety filtering, and related tasks, spanning 2002--2024, we document that these ethical challenges are almost entirely omitted from discussion in most research papers using nude images. 
We find that computer science research papers routinely engage in the nonconsensual collection and dissemination of nude images by scraping images extensively from the Internet, publishing those images as examples in research papers, and including them in publicly disseminated nudity detection models and datasets. In this paper, we identify and quantify common harmful data collection, distribution, and research practices.

Our goal is not to further stigmatize nudity and sexuality. We recognize that there are legitimate research needs for datasets of nude images. Rather, we argue that \textbf{research on nudity detection and related tasks must stop nonconsensually collecting and distributing nude images in order to avoid engaging in image-based sexual abuse.} We believe there are alternative paths forward that better embody our community's ethical standards and respect the dignity of all people.

%% file: sections/2_background.tex
\section{Related Work} 
Our work focuses on the nonconsensual collection and use of nude images in research, but issues of consent with regards to data collection are not unique to nudity. Other scholars~\cite{gebru_beyond_2024, jo_lessons_2020, birhane_large_2021, scheuerman_human_2023, gebru_datasheets_2021, paullada_data_2021} have critiqued the widespread practice of scraping data without consent to amass datasets for training and evaluating machine learning models. 
Previous work~\cite{scheuerman_human_2023} found that among 125 machine learning datasets, only the authors of two datasets specifically asked and received consent from their subjects. Subjects are unlikely to be aware of what their data are being used for, and even if they are, it is common for their data to be used beyond what they initially consented to~\cite{andreotta2022ai}. Furthermore, there are often no avenues for subjects to demand their data be removed from datasets, nor is this possible when datasets are shared and re-shared~\cite{scheuerman_human_2023, luccioni2022framework}. 
Our work details how these same norms, when applied to nude images, perpetrate image-based sexual abuse. However, the harms that we elaborate on in this paper are symptom of ``a culture where the appropriation of images of real people as raw material free for the taking has come be to perceived as the norm''~\cite{birhane_large_2021}. 

We are not the first to address the harms of nonconsensually using nude images in machine learning research.
Prior work has documented the existence of nonconsensually shared nude images in large image datasets (e.g., ImageNet~\cite{birhane_large_2021}, Moments in Time dataset~\cite{everest_pipkin_lacework_2020}, LAION-400M~\cite{birhane2021multimodal}) that are used for a wide variety of machine learning tasks. 
Most notably, researchers in 2016 conducted a review of 102 computer vision papers that focused on ``pornography filtering.'' Their work surfaced the gender and sexual politics that are embedded in how computer vision researchers discuss and conceptualize their research (e.g., the common usage of female genitalia as a proxy for detecting ``pornographic images''~\cite{gehl_training_2017}). Our work, nearly a decade later, observes similar trends but focuses on the harms of research practices around the collection, distribution, and handling of datasets that are described to primarily contain nude imagery.

%% file: sections/3_methods.tex
\section{Methods}
\label{sec:methods}

We systematically analyzed how researchers describe and use datasets containing nude images in computer science research. \edit{To do this,} we collected computer science publications based on a set of search terms, manually annotated papers to extract relevant information for analysis, and then conducted both quantitative and qualitative analyses. In total, we collected \edit{\nPapers}~papers after applying our inclusion criteria, \edit{which we manually annotated using a list of pre-defined questions}. Additional details about our methodological process are captured in Appendix~\ref{appendix:methodology}.

\textbf{Data Collection.}
We created an initial set of keywords based on exploratory review of highly-cited papers in nudity detection. We identified commonly used datasets and recurring task names associated with the use of datasets containing nude images (e.g., content filtering, adult image classification). Our final set of search terms can be found in Appendix~\ref{appendix:search-terms}.
We then used our search terms in Google Scholar, \edit{which searches the full text of manuscripts,} and collected 1204~papers for further review.
Our intention was to capture research whose technical tasks rely on nude image datasets. Our data collection therefore does not exhaustively cover all publications that use datasets containing nude images.

\textbf{Inclusion Criteria.}
Our keyword search procedure returned 1204~papers. 
We focused our inquiry on 
(1) computer science publications that (2) use a dataset containing real (non-generated) nude images. We limited our analysis to computer science venues to allow for clearer comparisons in dataset framing, usage, and justification across papers that share technical assumptions, audiences, and publication norms. While the majority of these venues were related to AI/ML, 10 were related to security/privacy. However, almost all of the papers were centered on techniques from AI/ML.
We filtered papers in our dataset to only publication venues indexed by DBLP~\cite{dblp}, a database that catalogs computer science literature. The DBLP team provided us with a list of all of the venues they index.
This filtering left us with a set of 379~papers. 
To address criteria (2), our team read through each paper and manually determined if the paper used any real nude images.

\textbf{Dataset Descriptions.}
After applying our inclusion criteria, we arrived at a final set of \edit{\nPapers}~papers and found that there is urgency in addressing the issue of nonconsensual nude image collection and distribution in research. 
While our set of \edit{\nPapers{}} included papers that were published between 2002-2024, the majority \edit{(76)} were published in \edit{2019} or later (see Figure~\ref{fig:papers-by-year-range}).
The use of datasets containing nude images is occurring across a multitude of venues. In total, the papers spanned 110 venues, including \edit{eight} papers at six A* venues (based on CORE rankings~\cite{core-rankings}) such as CVPR and AAAI (see Appendix~\ref{appendix:methodology-annotation} for full list). 
Furthermore, many papers were published through prominent publishing organizations such as IEEE (\edit{52} papers) and the ACM (\edit{10} papers). The papers in our inclusion set had collectively received \edit{5,846} citations at the time of collection (Fall 2024), with the median number of citations being 10. This demonstrates that machine learning tasks involving the use of nude datasets are an active area of research.

We observed that nude images are used by researchers around the world. In total, the papers were authored by researchers at \edit{over 200} institutions across \edit{42} countries, including the United States where 12 of the 13 institutions were R1 universities, such as Carnegie Mellon, Georgia Tech, Northeastern University, and UC Berkeley.

\textbf{Annotation.} 
\edit{\textbf{We manually annotated all \nPapers{} papers} that met our inclusion criteria using a list of pre-defined questions (see Appendix~\ref{appendix:methodology-questions}). Each paper was annotated for (1) key details about the dataset (e.g., how many nude images were collected and used), (2) information about any published example images, and (3) research practices around data handling. 
We also defined a set of supplemental questions to investigate the framing and stated goals of research in our inclusion set (available in Appendix~\ref{appendix:methodology-additional-questions}). We captured this information using direct quotes from the papers.
We randomly sampled and independently annotated papers from our inclusion set with these additional questions until we collectively reached thematic saturation~\cite{guest2006many}. This process included five papers published at A* venues.
Since A* venues may be indicative of (and also set) norms for the rest of the research community, we identified and included 3 additional papers in our dataset that were published at A* venues. In total, \textbf{\nFullyAnnotated~papers} were annotated with the supplemental framing questions.}

\textbf{Data Analysis.}
We analyzed the results from each annotation question using methods according to the data type. Most of our annotation questions required numerical or categorical responses (e.g., ``How many nude images are in the dataset?'') that we quantitatively analyzed to get descriptive statistics. To answer the  annotation questions that required qualitative responses (e.g., ``Does the paper discuss a social goal?''), we directly copied quotes from the papers. We then analyzed the quotes through an inductive coding process that is discussed in greater detail in Appendix~\ref{appendix:methodology-analysis}.

\textbf{Ethical Considerations.}
Due to concerns about the inclusion of sample images in papers and some of the research practices we describe in our findings, we have taken measures to reduce the discoverability of the papers within our corpus.\footnote{We will make the dataset of papers available for researchers upon request on a case by case basis.} We intentionally do not share links to public datasets found during our analysis (though we make exceptions for well-known datasets, such as NudeNet). Instead of referencing specific papers, we have assigned each paper a random numerical ID. Papers published in A* venues are designated as PaperID$^{A*}$.
We are in the process of responsibly disclosing the papers that contain uncensored nude images (with faces) to relevant publishers (e.g., IEEE, ACM). 
More details about this process and researcher safety are included in Appendix~\ref{appendix:methodology-ethics}.

%% file: sections/4_results.tex
\section{Data Collection and Distribution Practices}
\label{sec:data-collection}
Through our analysis, we observed a pattern of widespread nude image collection without the consent of the subjects depicted.
Across the \edit{150} papers in our annotated dataset, \edit{126} unique datasets were used by researchers \edit{(including both existing and newly created datasets)}. 
\edit{In total, over 8 million images were collected and used.}
The largest dataset (Paper1052) contained 5,002,000 nude images and the smallest dataset consisted of \edit{50} nude videos (Paper252). 
While \edit{39 papers exclusively used previously collected} datasets, \edit{almost 2 million images} were newly collected by \edit{98} papers that created their own datasets. \edit{Thirteen papers did not specify any details about their dataset.}

\subsection{Using nude images nonconsensually in research is harmful}

\paragraph{Subjects' lack of consent is unacknowledged.}
\edit{Out of the \nFullyAnnotated{} papers that we annotated using our full set of questions,} we did not find any papers that discussed (the lack of) consent from image subjects. Furthermore, despite the real risk of emotional and physical harm that comes from redistributed nude content~\cite{henry_image-based_2019}, none of the papers we analyzed explicitly considered the image subjects' safety in their decision to conduct the study---despite using this data to purportedly enhance the safety of the Internet (see Section~\ref{subsect:research-objectives}).
This lack of consideration for subjects is well captured in an essay by cultural worker Livia Foldes on the nude images published through \edit{the most popular pretrained nudity classifier,} NudeNet, which noted that, ``the agency of the people captured in [the dataset] has been so thoroughly denied that their consent, or lack thereof, was never mentioned by the researchers who used their intimate photographs to train an algorithm''~\cite{livia_foldes_nsfw_2024}.

Regardless of whether the content being scraped was originally created and/or distributed consensually, it is being collected nonconsensually by researchers for use cases that are unbeknownst to the data subjects. Studies on other social media platforms have found that users are largely unaware that their ``public'' posts could be used for research, and preferred to be asked for consent before their data was used~\cite{fiesler2018participant}.
Similarly, those who have voluntarily shared their nude images online likely did not anticipate the possibility of having them collected, stored, and redistributed by researchers. 
\edit{Even when a platform has formally authorized data access for research use (e.g., through an API or terms of service),} \edit{researchers should not use a platform's terms of service as the arbiter of ethical data use and instead consider the safety and dignity of the data subjects across the benefits and context of the research.}

\paragraph{Using commercial content is still harmful.} To avoid collecting content that was nonconsensually created or shared, some researchers instead scraped from adult content platforms. \edit{In total, 18 papers stated that they collected data from adult content platforms while another 11} used a particular scraping tool that contained almost 4000 links to content on adult content sites. 
However, scraping data from adult content platforms is still a violation.
Sex workers are often excluded from discussions around IBSA, yet they face the same mental health harms from experiencing IBSA, as well as the risk of being outed as a sex worker \edit{and the loss of livelihood from stolen content,} among other threats~\cite{scarlett_redman_visual_2022, qin_did_2024, west_image-based_2024}. 
Furthermore, while large adult content platforms have more robust content moderation practices, images shared for the purposes of perpetrating IBSA commonly end up on smaller adult content platforms, some of which have no mechanisms for victim-survivors to seek content removal~\cite{huber_non-consensual_2024}.

\paragraph{\edit{Nude images are collected in excess.}}
The lack of consideration of image subjects is further evident in the \edit{excess} collection of new images. 
As mentioned, \edit{98} of the papers in our annotated corpus created new datasets. This decision is rarely supported by additional reasoning. When justification was provided, it was short and vague.
For example, Paper862 created a new dataset consisting of 18,000 images to have ``diverse and challenging images.'' Paper789 chose to develop a new dataset of nearly 38,000 images, claiming that ``there is no current dataset accessible for the classification of multiple body parts nudity.'' 
\edit{Meanwhile, Paper1100$^{A*}$ collected ``professional and amateur pornography'' that was ``intentionally recent'' to avoid overlap with existing data}. 

Norms around collecting ``public'' data without consent are extractive, regardless of the type of data collected~\cite{mittal2024responsible, jo_lessons_2020}. 
However, the same practices applied to nude content can lead to even further harm and legal risk \edit{for researchers, necessitating higher standards for care and caution}.

\subsection{Researchers are amassing datasets of IBSA}

Researchers are further amplifying harm by collecting and using abusive content. \edit{The vast majority (98)} of papers we annotated created new datasets.

\paragraph{\edit{Nonconsensual intimate content is difficult to remove from the Internet.}}
Scraping nude images from the Internet, especially social media platforms, will almost inevitably result in collecting some that were nonconsensually created or uploaded~\cite{henry_image-based_2019}. \edit{Perpetrators often nonconsensually distribute intimate content to social media platforms, online forums, and adult content sites. }
Once shared, the process of content removal is labor-intensive and can be re-traumatizing for victim-survivors. It may takes weeks for platforms to respond to their requests, if they respond at all~\cite{huber_non-consensual_2024}. In the meantime, victim-survivors repeatedly experience harm by spending hours submitting additional content removal requests and checking if the content is still public. \edit{Unmoderated adult content platforms may not offer any meaningful process for content removal.} By collecting and using this data, researchers inadvertently increase its visibility and perpetrate harm by creating new avenues for dissemination.

\paragraph{Abuse material is incidentally collected.} Of the \edit{98} papers that created new datasets, \edit{the vast majority (60) did not mention a specific data source}.

Those that did reported collecting data from Reddit (Paper998, Paper802, \edit{Paper1100$^{A*}$}), search engines including Google, Bing, and DuckDuckGo (Paper88, \edit{Paper777$^{A*}$, Paper815}, Paper976), 
\edit{Tumblr (Paper795)}, ``a large image search index (of a few billion images in size)'' (Paper1038$^{A*}$), ``public social networks or image sharing sites'' (Paper 862), and 4chan (Paper773$^{A*}$). Such platforms are also \edit{commonly used for disseminating} nonconsensual intimate content, \edit{sometimes in ways that victim-survivors are unaware of. Once discovered, victim-survivors often encounter difficulties with content removal}~\cite{huber_non-consensual_2024}. 
Researchers might therefore be unintentionally collecting abusive content.

\paragraph{Abuse material is intentionally collected.} In other instances, researchers intentionally included abusive content. Despite acknowledging that ``it is illegal to possess and distribute [upskirt] images,'' Paper887 collected 1,637 upskirt images from ``the Internet.'' Meanwhile Paper307 constructed a dataset that they described as being extracted from ``hidden or self cameras.'' 
Paper6$^{A*}$ uses a publicly-accessible dataset of URLs for 1,300,000 ``NSFW images.'' Upon inspecting the URLs, we found that they pointed to images on specific subreddits.
One of the folders
intentionally included content depicting sexual violence, collected from two subreddits that have since been banned from the platform (``r/StruggleFucking'' and ``r/rape\_roleplay'').

\edit{Eighteen} papers specifically mentioned collecting data from adult content platforms.
In particular, Paper873 described collecting ``unprofessional porn'' that was ``taken by unprofessional photographer including both
nudity and sexual behaviour. The unprofessional porn images usually contain complex backgrounds and the image quality is poor.'' Based on the description alone, it is likely that the authors either collected stolen content from small-scale creators or collected content that was nonconsensually created (e.g., through a hidden camera). \edit{Similarly, Paper936 collected ``amateur pornographic images.'' As previously mentioned, it is particularly challenging to remove nonconsensually distributed content from small, unmoderated adult sites since some do not provide any means of reporting content.}

\edit{Our corpus also included papers (Paper1090, Paper806, Paper1083) that used child abuse material by partnering with law enforcement directly. These datasets were held by law enforcement, rather than the researchers, and had strict data handling protocols that prevented the researchers from ever viewing the content. This is a positive indication that this type of research is possible to do responsibly. However, not all papers did this.}
\edit{One paper (Paper815) described CSAM detection as a goal of their research but did not partner with law enforcement. Furthermore, they alluded to} collecting and reusing nude images of children and minors with keywords such as ``girls, boys, teenager...'' in conjunction with keywords related to sexual acts (``licking, lick, sucking, suck, blowjob, fellatio''). This likely constitutes CSAM and is illegal in most jurisdictions~\cite{icmec2016}.

\paragraph{Data sources are largely undisclosed.}
We emphasize that the observations made above are based on a small number of data source disclosures (or through further investigation of datasets and third-party tools). Of the \edit{98} papers that created new datasets, many did not state a data source \edit{(19)} \edit{or included vague references to the ``Internet'' or ``websites'' with no specified source (60)}.
Of the \edit{38} papers that mentioned a specific online source, most descriptions were short and vague (e.g.,``samples from public social networks or image sharing sites'' (Paper862)).
By purposely obscuring the data sources or neglecting to include these details, researchers are evading accountability. 
Therefore, our observations are an under-reporting of the full extent of harm and the potential illegality of images that have already been collected within our corpus.

\subsection{Images are distributed without consent via publication, annotation, and open science}
\label{subsect:distribution-channels}

We identify three primary pathways through which ML researchers further distribute nude images without the image subject's consent: publishing example images in research papers, sharing with annotators, and disseminating image datasets and models trained on those datasets. \edit{We emphasize that this further distribution of nude images \textit{without consent} fundamentally constitutes IBSA.}

\textbf{Publishing example images.} A vast majority \edit{(87)} of papers in our corpus, including those published in A* venues (4), embedded example nude images in their papers. \edit{Nine papers did not censor the images at all and 28 papers only censored  (e.g., applied a blur effect) the bodies of those depicted, while leaving faces visible.}
Over \edit{816} 
images were included across \edit{87} papers.
Paper898 contained over 40 images (faces censored) while Paper815 contained over 10 images that were completely uncensored.
Six papers, including Paper6$^{A*}$, included example images of explicit sexual acts. Paper887 published 21 upskirt images, despite acknowledging that such images are illegal to disseminate in some jurisdictions.

One of the primary challenges that victim-survivors of IBSA face is stopping further dissemination of nonconsensually created and/or shared content~\cite{ward_revenge_2022, huber_non-consensual_2024}. The distribution of nude images as examples in academic papers contributes to this challenge. Example nude images are unnecessary for readers to understand nudity as a concept; text descriptions or artistic depictions would suffice.

\textbf{Sharing with annotators.} 
Many papers \edit{(44)} described the need for manually labeling nude images, including annotating ``centers of private parts'' (Paper883, Paper1066, Paper1065) and specific sexual acts such as ``Vaginal Penetration, Anal Penetration, BDSM, Bestiality'' (Paper815). In total, researchers mentioned annotating \edit{over 800,000}  
images across \edit{44} papers. 
Many of these papers were opaque about who performed annotations. For example, Paper789 had only two authors but ``eighteen human annotators.'' Only four papers explicitly note that the researchers outsourced annotation work, such as Paper873, which ``employ[ed] six person[s]'' to annotate the dataset, \edit{and Paper978, which ``invited 10 students in our laboratory to manually filter out the images,'' which led to a dataset of ``nearly 150,000 pornographic images and about 500,000 normal images.''}
\edit{Three of the four papers that outsourced annotation} provide no details of annotation policies, tools involved in the annotation process, protections against dissemination of the nude images by annotators, or \edit{protections for annotators against viewing abusive content (given that some researchers collected abusive content)}.

\textbf{Releasing datasets.} Nude datasets serve as a concentrated repository of nonconsensually collected and distributed intimate content, making it more difficult for image subjects to seek removal of material depicting them from the Internet. Although there are norms within the ML community towards publicly sharing datasets, public accessibility is not appropriate in this context. 
We found \edit{three} papers, including Paper777$^{A*}$, that made their datasets publicly available. Nude datasets from Paper777$^{A*}$ and Paper862 are still publicly accessible at the time of this publication submission through HuggingFace and figshare.
Furthermore, \edit{three} papers made nude datasets available upon request.
\edit{While three papers mentioned deciding not to distribute their dataset, decisions to limit or withhold access were not necessarily motivated by concerns about the privacy of those depicted.}
Paper998 expressed desire to make data public but was thwarted by existing laws: ``Despite our every effort to make the dataset content and data acquisition process as transparent as possible, due to the nature of data and applicable laws the dataset cannot be shared publicly.''
\edit{Furthermore, Paper845 made the decision not to share the data because, ``the images used in this paper... likely contain copyrighted content,'' rather than as an abuse or privacy consideration.}

\section{Research Practices}

Established research ethics require that researchers decide what research practices are morally acceptable to achieve their aims. Prior work on ethics in computer security~\cite{kohno_ethical_2023} highlights two core ethical frameworks that researchers can use to reason about their decision to engage in behavior that harms individuals. 
Here, we leverage both frameworks---consequentialist and deontological ethics---as lenses through which to consider the justifications for the research practices we observed. Further, we consider how existing mitigations against unethical research practice---institutional review boards and publishing-venue-mandated ethics statements---have failed to guard against harm.

\subsection{Research objectives do not justify the means}

\label{subsect:research-objectives}
Consequentialist ethics centers a tradeoff between the harms and the benefits of a decision. To justify such a tradeoff, the benefits of the decision must be clear \textit{and} outweigh the harms.

\edit{Of the {74} papers we annotated for framing,} the vast majority \edit{(51)} of those that specified any motivation \edit{(59)} reported their aim as reducing the prevalence of pornography on the Internet. Most of these papers described the benefit of their work as combating the moral and societal ills of pornography without providing concrete examples or citations. For example, papers assert that ``Internet pornography content affects many people's life especially adolescents and creates many social problems and moral issues'' (Paper1079) and ``The amount of digital pornographic content over the Internet grows daily and accessing such a content has become increasingly easier. Hence, there is a real need for mechanisms that can protect particularly-vulnerable audiences...'' (Paper821). Yet, pornography is legal in a large number of jurisdictions, and scientific evidence about its harmfulness to adults is inconclusive at best, with recent meta-reviews suggesting that experiences of pornographic use as problematic are ``actually more related to [individuals'] interpretations of that use rather than the use itself. Specifically, religious
qualms...and moral disapproval of'' their own use of pornography~\cite{grubbs2021pornography}. 

Further, not all nude images on the Internet are sexual or pornographic. People have a wide range of purposes for posting and consuming sexual and/or nude content, including to explore their sexuality, seek positive body validation from friends, or receive sexual education~\cite{paasonen_about_2023, geeng_usable_2020, qin_did_2024, waldman_law_2019, macdowall_sexting_2022}.
Indeed, \edit{condemning} sexual content that is not abusive may in fact cause harm: prior work suggests that the censorship of sex and sexual speech on the Internet (i) increases stigma, therefore harming victim-survivors of image-based and other forms of sexual abuse, and (ii) decreases general access to sexual and reproductive health resources, which in turn plays a role in preventing in-person and online sexual abuse of both adults and children~\cite{walker2017systematic,henry_image-based_2022, women2018international,stardust_indie_2024}.
Furthermore, there is hypocrisy in claiming the moral need to remove nude and sexual content from the Internet while creating repositories of such content and aiding their spread through the distribution mechanisms discussed in Section~\ref{subsect:distribution-channels}.

While more than half of the papers \edit{we annotated for framing} purported to protect children from the harms of pornography, only \edit{six} papers mentioned the issue of CSAM. Even fewer specifically worked on technical solutions to address this issue.
This has echoes of the classic use of child protection to induce moral panic~\cite{marwick2008catch} and justify censorship, rather than an articulation of work on a substantive societal problem that could justify the consequences of harmful research practice.

\subsection{Subjects are dehumanized}
The other ethical framework highlighted by Kohno et al.~\cite{kohno_ethical_2023}, deontological ethics, at a high level suggests that ``we have a duty to treat all other human beings... as `ends and never purely as means.''' Under such a framework, there could be no justification that any societal benefit (an end) justifies using an individual depicted in an image as a ``means'' by using that person's nude images without consent (violating multiple of their rights: to privacy, autonomy, etc.). Given the language several papers use to refer to nude images, however, we hypothesize that the researchers may have so dehumanized the people depicted in the content that they have forgotten that they are people 
worthy of respect and protection. For example, papers described nude images and/or sexual content as ``obscene'' (Paper1079, Paper1095, Paper878, Paper39), ``virus'' (Paper1066), ``offensive'' (Paper897, Paper899), ``harmful'' (Paper1184$^{A*}$, Paper919) and ``inappropriate'' (Paper1187 and Paper6$^{A*}$). %

\subsection{Ethical considerations are overlooked} 

We specifically annotated papers in our corpus for discussions of ethical or privacy concerns and only found four papers (Paper6$^{A*}$, Paper773$^{A*}$, Paper853, Paper733) that made any mention of either.
In these handful of instances, three papers focused on data access. Protective measures included limiting access to one individual (Paper853), not making data public (Paper6$^{A*}$), and not using crowdsourced/third-party annotators (Paper773$^{A*}$) (though the concern in this paper was to avoid exposing annotators to ``disturbing and unsafe'' images, rather than protecting the image subjects).

\edit{Out of the 150} papers we annotated, \edit{only eight} discussed data security. \edit{For example, Paper733 and Paper806 mentioned using designated servers. Four papers (Paper733, Paper853, Paper724, Paper773$^{A*}$) limited dataset access to only authors of the papers to prevent ``exposing harmful content to other people'' (Paper724)}.
\edit{Even then, none of the papers} mentioned any deletion plans for the nude images they used. We urge researchers who have already collected datasets of nude images to remove them from any public repositories they are shared to and to securely delete images that are no longer being used.

Only \edit{two} of the papers we annotated, \edit{Paper724} and Paper1108$^{A*}$, mentioned seeking (and receiving) IRB approval. Institutional review boards such as IRB or ERB offices must consider the potential harms and legality of collecting nude images from online spaces. Such boards should require researchers to articulate their data sources, construct detailed data security plans, and disclose any third-party data sharing (either direct sharing with other individuals and/or with third-party tools).

\subsection{Classification boundaries lack thoughtful definition}

\textbf{Classification objectives are largely undefined.} 
\edit{All but one paper we annotated for framing} 
described their technical purpose as detecting nude, ``adult,'' and/or ``pornographic'' images.
Despite making claims such as, ``exposure to the sea of pornography can lead to many social problems, including cyber-sex [addiction]. It is now an urgently necessary task to prevent people, especially children from accessing this type of harmful material'' (Paper823), the papers we analyzed commonly did not articulate what they were trying to remove. 
\edit{More than} half (39) of the \edit{fully annotated} papers did not provide any specific criteria for classification. The majority of criteria (35) related to identifying body parts ranging from specific mentions of ``anuses, female breasts, female genitals, male genitals'' (Paper1108$^{A*}$) to vague references (e.g., ``private body parts'' (Paper1184$^{A*}$), ``certain parts of the body'' (Paper862)). 
The lack of specificity not only makes it difficult to understand the accuracy of model evaluation, but suggests they may capture entirely unrelated content.
The lack of clear classification criteria establishes a norm that disregards censorship as an outcome.

\edit{Broad and underspecified classification criteria are also misaligned with the content moderation needs of online platforms, which must carefully navigate the balance between moderation and censorship~\cite{are_shadowban_2022, gillespie_custodians_2018, tiidenberg_sex_2020, paasonen_nsfw_2019, tiidenberg_sex_2021}.
Many papers that we fully annotated justified collecting and using nude datasets by suggesting that their work could be adopted by online platforms. Yet, what qualifies as nudity, in the context of content moderation, is nuanced, contested, and continually negotiated between users and large platforms. For example, Meta has revised their policy multiple times in response to user concerns about censorship and inequitable enforcement \cite{meta_transparency_2025}. As a result, instances of nudity/sexual content related to health, education, and artistic expression are now excluded from moderation. In contrast, none of the fully annotated papers we reviewed accounted for context-specific exceptions, suggesting that researchers' criteria are disconnected from how content moderation occurs in practice.}

\textbf{Classification is fundamentally difficult.}
The delineation of what is nude/non-nude or sexual/not sexual is inherently subjective. There is a long history of debate in legal literature regarding these definitions~\cite{arthur2018problems}, yet only one paper (Paper799) in our \edit{fully annotated} corpus discussed defining the classification threshold as a limitation of their work. \edit{More than} half (39) of the \edit{fully annotated} papers lacked any specific criteria for classifying content as nude (or sexual) or not nude (or non-sexual), treating the threshold as well-understood and objective (e.g., ``each image contains an obvious instance of pornography'' (Paper1090)).
In other cases, researchers made arbitrary classifications without justification or definitions, such as Paper799 which scored the ``severity'' of content (from less to more severe): ``Female breast 1, Female buttock 2, male buttock, 2, Female genitalia posing 2, Female genitalia sexually active 3, Male genitalia 3, Sex toys 3, Coitus 4, Anal 4.''

These classification and definitional decisions have large-scale societal consequences.
Nudity classification algorithms disproportionately rate images of women as sexually suggestive~\cite{riccio2024exposed}. For example, images of women athletes are more likely to be flagged as depicting nudity or being sexually suggestive than similar images of athletes who are men~\cite{gianluca_mauro_there_2023}. Indeed, a prior analysis of computer vision literature well identified the implicit assumption ``that pornography is limited to images of naked women; that sexuality is largely comprised of men looking at naked women''~\cite{gehl_training_2017}. 
While we cannot ascertain the gender of a subject based on an image, we observed that the example images \edit{in papers aligned with their analysis}.
\edit{We urge researchers to directly engage with the inherent subjectivity in classification and reflect on how their positionalities (e.g. personal values, politics, experiences, cultural values) affect the classification objectives they form~\cite{cambo_model_2022, scheuerman_how_2025}.}
Furthermore, by creating a binary classifications of whether an image is ``pornographic'' or not, ``voluntarily shared nudes, commercial porn, sex work, human trafficking, queer activism, NSFW fanart and paedophilia will converge, unhelpfully, into a muddy pit of obscenity''~\cite{tiidenberg_sex_2021}.

%% file: sections/5_discussion.tex
\section{Discussion}
Although datasets containing nude images have been used for well over two decades in computer science research, there has been no guidance on their usage aside from a relatively recent ban on the use of one specific image: a headshot collected nonconsensually from a Playboy centerfold in 1972~\cite{mulvin2021proxies}. We observed virtually no difference in terms of how researchers handled nude images compared to other image and vision datasets. Despite dealing with highly sensitive data, papers largely make no effort to describe data storage, access policies, and security practices to ensure that the data is safe.
More troubling, many papers themselves further disseminate nonconsensual images, and offer no---or very weak---justification for the harms of the research process. Here, we zoom out and discuss responses the computing community can take to avoid this harm in the future.

\subsection{Publishing venues should step up to regulate and set norms}
Frontier venues and major publishers such as NeurIPS, ACM, and IEEE set standards for the entire computing community---both in terms of scientific rigor as well as ethical norms. Over the last several years, most frontier venues have established research ethics guidelines that submitted papers are supposed to adhere to, and are increasingly defining the processes that reviewers should use to evaluate and potentially reject studies based on these guidelines. Existing principles, such as those around privacy~\cite{neurips_ethics,icml_ethics,cvpr_ethics,acm_ethics,usenix_ethics}, consent~\cite{neurips_ethics,acm_ethics,cvpr_ethics,usenix_ethics}, and minimizing harm~\cite{neurips_ethics,icml_ethics,cvpr_ethics,acm_ethics,usenix_ethics}, are sufficient to cover the practices we highlight here.

Of the \edit{eight} papers published at ${A*}$ venues that we annotated, five were published in 2023 or later, by which time every venue had ethics guidelines already in place.
This suggests that, while existing ethics guidelines cover principles that should regulate the use of nude images in research~\cite{arya2024misuse}, reviewers are not raising concerns sufficiently to trigger a change in practice or additional discussion in the text of the papers. 
We urge reviewers to take additional care when reviewing papers that use nude images, especially related to data sources, dataset handling, appropriateness of the use cases, and the presence of example images.

Furthermore, while existing research ethics guidelines should be broad enough to cover harmful practices around nude image collection and use, publishing venues should take a stronger step to prevent harm by banning (and removing) the use of nude images in published manuscripts \textit{where the subject is identifiable}. 
This, however, should not be taken as guidance for the wholesale banning of nude imagery in research publications, \edit{which may be relevant to the research in some cases}. 

\subsection{Moving forward}

As a research community, if we accept that there exist tasks for which highly sensitive and potentially harmful data are necessary, it is imperative that we also forge pathways to the ethical creation and handling of such data.

\edit{\textbf{Research with a clear purpose.} We call for researchers to carefully evaluate and clearly articulate the benefits of research that necessitates the collection and use of nude datasets. If the intention is for research to be adopted by law enforcement or industry, we urge researchers to partner directly with relevant stakeholders. This will ensure that design decisions are aligned with their real-world needs.
Building partnerships will also enable researchers to leverage existing data that stakeholders may already have access to (e.g., previously flagged content). }

\edit{\textbf{Opportunities and risks of AI-generated data.}} While this study examined papers using datasets of real people, our keyword search turned up many studies that were already using \edit{AI-}generated nude images for various tasks. \edit{While generative AI might have the potential to be} a less harmful strategy for creating a dataset of nude images, it is important to note that generative AI is not a harm-less approach. 
\edit{These models are most likely trained using nonconsensually collected and/or created images~\cite{thiel2023identifying}.
Researchers should take several things into consideration when deciding whether generated nude images are appropriate for their task. The risks and opportunities of this approach are still evolving. Indeed, members of our own research team hold differing stances on how to balance the possibility of harm reduction through AI-generated data and the possibility of furthering harm through the use of imperfect tools.}

\edit{First, as discussed above, we encourage researchers to engage in their papers with how the goals of the research task outweigh the real costs of using generative models trained on nonconsensually collected and created content.
Second, researchers should explicitly acknowledge in their manuscripts that current generative models are trained on the same data that researchers are trying to avoid using. Finally, researchers must think critically about the images they generate. For example,} though unlikely, an image used in the training set may be re-produced as output. We also do not yet know how likely it is for an output to closely resemble the likeness of a living person regardless of whether their image was used in training. More research is needed to assess these risks.
We also note that generated images may not be suited for all research tasks.
As such, generating images does not negate the need to develop a new, ethical model of data governance for real nude images.

\textbf{The need for a new model of data governance.} We call for a new model of data governance that centers ongoing, informed consent. Specifically, we propose a participatorily-governed data trust of nude and sexual images that are consensually collected for research purposes. 

Prior work has shown that people are willing to donate highly sensitive data to research for altruistic reasons, and are more willing if they understand the purpose of the research~\cite{skatova2019psychology, razi2022datadonation}.
While further research is needed to understand the circumstances under which people would be willing to donate nude and sexual imagery to science, their willingness will also depend on their perceptions of the value of the research and the trustworthiness of the data trust. Beyond a mechanism to ensure that participants fully understand what it means for their data to be used for research, participants should be able to specify which types of tasks they are willing to participate in. For example, a participant may be willing to donate data for use in nudity detection, but wish to be excluded from any generative models. 

While we believe such a data trust is feasible, it raises several challenges that will require future research. For example, a data trust will need careful measures to ensure that the donated data was created and donated consensually. One of the challenges we highlight above is that there is no way, from an image alone, to tell whether it was created consensually. Verifying consent will be critical for building a viable dataset. Furthermore, participatory governance by data subjects will require implementing complex access and use specifications and careful oversight of the ways that researchers use the data. 
A data trust would need careful data protection mechanisms, but its creation can include strong safety guarantees, for example by ensuring that a resulting generative model has not memorized them or that their likeness has not been generated.

\smallskip

Ultimately, while our research has uncovered many ways that the research community is in need of intervention on this problem, we believe there are pathways forward that better embody our ethical standards and protect the dignity of all people. 

%% file: sections/appendix.tex
\section{Methodology (Additional Details)}
\label{appendix:methodology}
\subsection{Data Collection}
\label{appendix:search-terms}
We used the following search terms to collect our initial set of 1204 papers. As we identified common existing datasets of nude images, we also searched on those dataset names. We elide the dataset names here.
\begin{itemize}
    \item ``adult image detection''
    \item NudeNet
    \item ``nudity detection''
    \item ``pornographic image detection''
    \item ``pornographic images'' + ``machine learning''
    \item ``safety filtering'' + ``machine learning''
\end{itemize}


We searched for our selected list of keywords on Google Scholar, with each keyword or combination of keywords in quotes. We collected all pages of the search results using a Selenium-based web scraper. To address CAPTCHAs~\cite{Else2018} and ensure representative data collection, we ran the data collection across multiple systems (and therefore IP addresses). 
We used \texttt{scholarly}~\cite{scholarly} to collect structured metadata, such as venue, year, and author names. 
We manually annotated the affiliations of all the authors as this information was not present in metadata retrieved by scholarly. Lastly, we de-duplicated the data to remove overlapping papers that were collected across multiple search terms.


We matched the venues in our dataset to those indexed by DBLP by using strict matching and then used \texttt{fuzzywuzzy}~\cite{fuzzywuzzy} 
to match the remaining unmatched venues. Our goal in using fuzzy matching was to maximize recall in identifying papers published in computer science venues, even when venue names were formatted differently (e.g., ``CVPR 2019'' vs. ``Proceedings of the IEEE Conference on Computer Vision and Pattern Recognition''). 
We selected token\_sort\_ratio as our matching algorithm because it handles word order differences effectively when comparing venue names. A match was considered valid when the similarity score was greater than or equal to 85. We then manually verified all fuzzy matches. This filtering resulted in a set of 379 papers that were published at DBLP-indexed venues. 

\subsection{Annotation} 
\label{appendix:methodology-annotation}
After inspecting each of the 379 papers and applying our inclusion criteria, we were left with \edit{150} total included papers. \edit{Since the papers in our dataset include sensitive content that is potentially collected nonconsensually, such as identifiably nude example images, we will only make our dataset available upon request to minimize traffic to papers that distribute images and datasets.}

We collectively reviewed an initial set of 20 papers to inform the development of our annotation framework. After discussing our initial observations, we drafted a preliminary set of annotation questions to guide systematic coding in the larger dataset. 
Our annotation questions are documented below and include the following topics: dataset details (e.g., data collection methodology, quantity of images in dataset), whether/how images from the dataset were published, the framing of the paper (e.g., its technical and social goals), and dataset handling (e.g., how researchers interacted with the dataset). 
Three of the authors used the questions to independently annotate the 20 papers. 
We then met as a team to discuss the process and made revisions to create a finalized set of annotation questions. 
We then \edit{manually annotated 150} papers that met the inclusion criteria. \edit{Among the 150 papers, 74 papers were annotated using supplemental questions concerning framing, social motivation, and terminology used.} Papers were divided among three of the authors for independent deductive annotation until thematic saturation was reached, as previously noted in Section~\ref{sec:methods}.

\edit{Eight} papers were published at six A* venues:
ACM International Conference on Multimedia (MM), Computer Vision and Pattern Recognition (CVPR),
Annual ACM Conference on Computer and Communications Security (CCS),
European Conference on Computer Vision (ECCV),
USENIX Security, 
AAAI Conference on Artificial Intelligence (AAAI).

\subsubsection{Annotation Questions}
\label{appendix:methodology-questions}

\begin{enumerate}

\item \textbf{Inclusion / exclusion check}\newline
      \emph{Answer Q1 only; complete the rest of the form \textbf{only} if the paper is included.}
      \begin{enumerate}
         \item Dataset contains real (non-generated) images that are only or primarily nude?\\
               $\square$ Yes \hspace{1em} $\square$ No
      \end{enumerate}


\item \textbf{Dataset details}
      \begin{enumerate}
         \item Does the paper create a new dataset or reuse a dataset from another paper? \\
               $\square$ Create \;
               $\square$ Reuse \;
               $\square$ Both \;
               $\square$ Other\,(\rule{2cm}{0.4pt})
         \item If the dataset is reused, what dataset is it / what paper is it from?
         \item If the dataset is new, where did they collect the images from? If they do not say, state ``unspecified''.
         \item If the dataset is new, do they make it publicly available?\\
               $\square$ Yes—freely online \;
               $\square$ Yes—by request \;
               $\square$ Not available \;
               $\square$ Unspecified \;
               $\square$~Other\,(\rule{2cm}{0.4pt})
         \item What does the paper use the dataset to do? Be specific -- e.g., train and test their classifier, fine-tune an existing model, etc.
         \item How many nude images are in the dataset?
         \item What do they describe about the characteristics of the dataset, if anything? (e.g., gender, age, ethnicity, image quality, lighting, etc)
         \item What term does the paper use for describing the contents of the dataset? E.g.: nudes, explicit images, pornographic images
         \item \emph{Optional} Are there any other interesting datasets that they are using? E.g., datasets of bikini photos, the I2P text prompt dataset, a dataset of generated nudes, etc.
      \end{enumerate}

\item \textbf{Published images in the paper}
      \begin{enumerate}
         \item Does the paper include example images from the dataset?\;%
               $\square$ Yes \; $\square$ No
         \item If yes, how many images? 
         \item If yes, how are the images shown?\\
               $\square$ Totally uncensored (body \& face visible)\\
               $\square$ Body parts censored, faces uncensored\\
               $\square$ Fully censored (body parts \& faces obscured/cropped)\\
               $\square$ Other\,(\rule{2cm}{0.4pt})
         \item \emph{Optional} What other comments or observations do you have about the images visible in the paper? E.g. observed gender distribution, explicitness, etc.
      \end{enumerate}

\item \textbf{Dataset handling \& researcher interaction}
      \begin{enumerate}
         \item Does the paper discuss data handling protocols for the data? E.g. storage procedures, deletion plans, etc. Use quotes if possible.
         \item Does the paper describe any researcher interaction with the dataset? E.g. preprocessing to crop, manually annotating or inspecting annotations, etc. Use quotes if possible.
         \item Do they name a funding source? Only specify if it's clear that the project was directly funded, not if they simply thank e.g. their university.
      \end{enumerate}

      \end{enumerate}

\subsubsection{Additional Questions (paper goals and framing)}
\label{appendix:methodology-additional-questions}

      \begin{enumerate}
         \item What is the technical application of the paper? E.g. nudity detection, internet filtering, etc.
         \item Does the paper define nudity? State ``no'' or paste in a quote with the definition.\\
         \item Who is the ultimate user of the tool, if stated? E.g. tech companies who want to detect nudity, ISPs looking to filter content, parents to install on their children's phones, etc. If not mentioned, state ``unspecified''. Use quotes if possible.
         \item If mentioned, does the paper discuss a social goal? E.g., protecting children from nudity, purifying the internet, minimizing porn use at work, etc. Use quotes if possible.
         \item Does the paper discuss any ethical considerations they made during the research? Use quotes if possible.
         \item What terms does the paper use for describing sensitive aspects of the images? E.g.: genitalia, private parts, female breasts...
\end{enumerate}

\subsection{Analysis}
\label{appendix:methodology-analysis}
First, the lead author read through annotations corresponding to questions that required qualitative analysis to develop initial codebooks per question.
Afterwards, two of the authors independently coded the same set of 15 annotations (20\%) for each of the seven questions using the initial codebooks. 
Disagreements were discussed, which led to further revision of the initial codebook.
The rest of the annotations were then divided equally between the two researchers to be independently coded with the revised, final codebooks. The final codebooks are provided below in Appendix~\ref{codebooks}.

\subsection{Ethical Considerations}
\label{appendix:methodology-ethics}
\edit{As we found that some papers distributed nude images by including identifiable examples (i.e., with faces visible) in the published manuscript, we decided to contact the relevant publishers. We see this as a form of responsible disclosure, and sought to determine whether it was possible to get some of the most egregious images removed from or censored in the manuscripts being hosted.}
\edit{When contacting publishers and their ethics committees, we emphasized that our concern is around the identifiability of image subjects and their lack of consent in having their images re-published in an archival database, rather than a complaint about the presence of nudity in research papers, which may be appropriate in some contexts.}
\edit{At the time of press, we have met with and/or reported images to 4 publishers. While internal review may have begun, no manuscripts have yet been modified.}

At the start of the project, all researchers were aware that this project would entail reading and annotating papers that may contain nonconsensually shared nude images. We discussed the risks of engaging in this research.
Annotators were reminded to take breaks and were provided access to a trained mental health clinician who they could contact at any time for support.
\begin{figure}[h!]
  \centering
  \includegraphics[width=\textwidth]{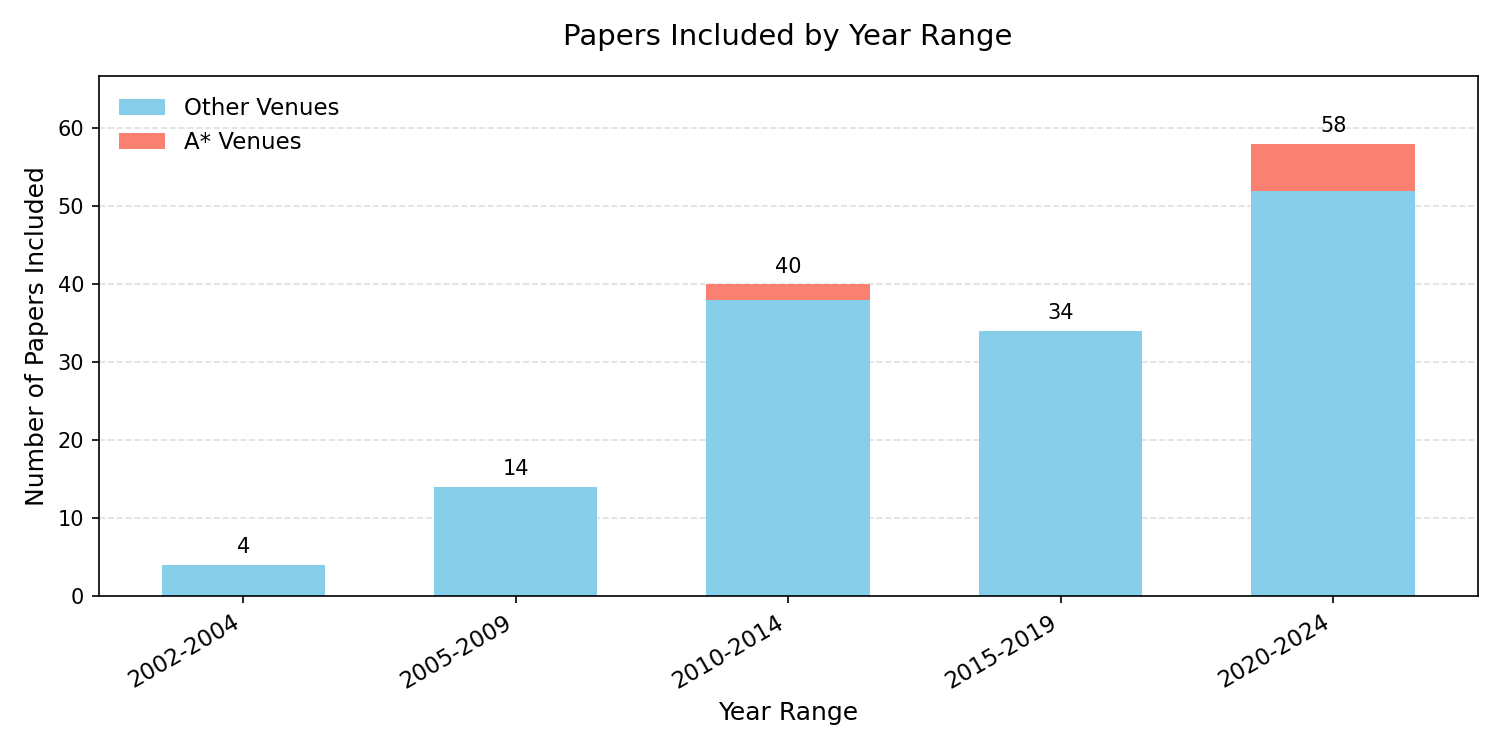}
  \caption{Number of papers included by 5-year range in the set of \edit{150} included papers. The stacked bars differentiate between A* papers (pink) and other venues (blue). Note that there were no papers from the years 2000 or 2001 in our dataset.}
  \label{fig:papers-by-year-range}
\end{figure}

\subsection{Codebooks}
\label{codebooks}

\begin{longtable}{|p{3.5cm}|p{10cm}|}
\hline
\textbf{Code} & \textbf{Definition} \\
\hline
\endhead

\multicolumn{2}{|p{13.5cm}|}{\textbf{Does the paper define nudity? State "no" or paste in a quote with the definition.}} \\
\hline
private\_parts & Definition includes exposure of at least one of the "private parts of the body": "female breasts", "vagina", "penis", and "buttocks" \\
\hline
naked & Definition includes naked bodies, without any clothes being present \\
\hline
skin & Definition describes large percentages of skin pixel or skin region \\
\hline
sex & Definition includes presence of sexual activities/scenes \\

\hline
female & Definition specifically mentions "female" body parts \\
\hline
male & Definition specifically mentions "male" body parts \\
\hline
unspecified & No definition of stated \\
\hline
\multicolumn{2}{|p{13.5cm}|}{\textbf{Who is the ultimate user of the tool, if stated?}} \\
\hline
companies & The tool is purportedly for companies (e.g., social media, video sharing, gaming) \\
\hline
parents & The tool is built for parents, mainly to protect their children from nudity online\\
\hline
children & The tool is built for children to protect themselves from nudity online (nudity ads, etc)\\
\hline
lea & The tool is built for Law Enforcement Agencies (LEA) to automatically detect CSAM and prosecute perpetrators who share CSAM\\
\hline
researchers & The tool is built to help researchers/AI practitioners to help with research or future work regarding nudity detection.\\
\hline
government & The tool purports to help the government identify content. \\
\hline
content\_mod & The tool is purports to help human content moderators \\
\hline
unspecified & No end user specified \\
\hline
\multicolumn{2}{|p{13.5cm}|}{\textbf{If mentioned, does the paper discuss a social goal of the paper? E.g., protecting children from nudity, purifying the internet, minimizing porn use at work, etc. Use quotes if possible.}} \\
\hline
protect\_minors & The social goal of the paper is to protect minors (children, teenagers, juveniles) \\
\hline
protect\_society & The social goal of the paper is to protect society 
\\
\hline
protect\_annotators & The social goal of the paper is to protect human content moderators or law enforcement agency staff from having to manually annotate content \\
\hline
purify\_internet & The social goal of the paper is to "purify the internet" \\
\hline
address\_csam & The social goal of the paper is to address CSAM \\
\hline
unspecified & The social goal of the paper is unspecified \\
\hline
\multicolumn{2}{|p{13.5cm}|}{\textbf{Does the paper discuss any ethical considerations they made during the research? Use quotes if possible.}} \\
\hline
not\_public & dataset cannot be made accessible \\
\hline
limited\_access & access limitations were made \\
\hline
beneficial & benefit surpasses harm \\
\hline
unspecified & no ethical considerations mentioned \\
\hline
other &  \\
\hline
\multicolumn{2}{|p{13.5cm}|}{\textbf{Does the paper describe any researcher interaction with the dataset? E.g. preprocessing to crop, manually annotating or inspecting annotations, etc. Use quotes if possible.}} \\
\hline
annotate\_nudes & manual annotation/labeling to determine nudes, sexual, and non-nude images and divide them into separate classes \\
\hline
annotate\_body & manual annotation/labeling to select specific body parts \\
\hline
annotate\_sex & manual annotation/labeling to select specific "sexual acts"/"positions"/"postures" \\
\hline
for\_children & manually removing images inappropriate for children \\
\hline
choose\_images & manually choosing nude images for created dataset/training set  \\
\hline
crop\_useless & manually cropping images to remove unnecessary parts/feature (e.g. background, logo. etc) \\
\hline
inspect & inspecting annotation, dataset flaws/mistakes, results/mistakes\\
\hline
none & no researcher interaction \\
\hline
unspecified & unspecified \\
\hline
\end{longtable}